\documentclass[journal]{IEEEtran}

\ifCLASSINFOpdf
  \usepackage[pdftex]{graphicx}
\else
  \usepackage[dvips]{graphicx}
\fi

\begin{document}
\title{Higher Harmonics Generation in Strongly Coupled Magnetized Two-Dimensional Yukawa Liquids}

\author{T.~Ott,
        M.~Bonitz,
        Z.~Donk{\'o},
        H.~K\"ahlert and~P.~Hartmann
\thanks{T. Ott, M. Bonitz, and H. K\"ahlert are with the Christian-Albrechts University of Kiel, Germany, Leibnizstr. 15, 24098 Kiel, Germany.}
\thanks{Z. Donk{\'o} and P. Hartmann are with the Research Institute for Solid State Physics and Optics, Hungarian Academy of Sciences, P.O. Box 49, H-1525 Budapest, Hungary. }
\thanks{Work was supported by the DFG via SFB-TR 24  (project A5 and A7), the HLRN Grant shp0006,  by OTKA via Grants No. K-77653 and No. PD-75113, and by a Janos Bolyai Research Grant of the HAS.}%
\thanks{Manuscript received April 19, 2005; revised January 11, 2007.}}

%

\maketitle

\begin{abstract}

The excitation spectra of two-dimensional strongly coupled (liquid-like) Yukawa systems under the influence of a magnetic field perpendicular to the plane are found to sustain additional high-frequency modes at multiples of the magnetoplasmon. These modes are reminiscent of the well-known Bernstein modes but show a number of important differences due to the strong coupling of the particles. 
\end{abstract}

\IEEEpeerreviewmaketitle

The two-dimensional Yukawa Plasma (2DYP) is being actively studied as a model for a diverse selection of physical systems, among them the grain-subsystem of dusty plasmas~[1,2]. The interaction potential for the 2DYP is $V(r) = Q^2/(4\pi\varepsilon_0)\times \exp(-r/\lambda_D)/r$, where $r$ is the particle distance and $Q$ their charge. In contrast to the widely investigated One-Component Plasma (OCP), the interaction of the 2DYP particles is screened. In the case of dusty plasmas, this screening is provided predominantly by the ion component of the plasma. The 2DYP is described by two parameters, the ratio of the Wigner-Seitz-radius and the Debye length, $\kappa=a/\lambda_D$, and the (OCP-)coupling parameter $\Gamma=Q^2/(4\pi\varepsilon_0 ak_BT)$, where $k_BT$ is the thermal energy of the system. 
Quasi-two-dimensional layers of dust grains are routinely generated in experiments, and efforts are currently underway to impose magnetic fields on these experimental systems. Furnishing the 2DYP with such a strong external magnetic field leads to the modified equations of motion [${\bf r}_i=(x_i, y_i)$]:
\begin{equation}
    m \ddot{{\bf r}}_i = Q\dot{{\bf r}}_i \times {\bf B}
- \frac{Q^2}{4\pi\varepsilon_0} \sum_{j\neq i}\left ( \nabla \frac{e^{-r/\lambda_D}}{r} \right )\Bigg \vert_{{\bf r}= {\bf r}_i - {\bf r}_j},  \, 
\label{eq:langevin}
\end{equation}
where $m$ is the particle mass and ${\bf B}= B {\bf e}_z$ is the magnetic field. To describe the magnetized 2DYP, one additional parameter is necessary, the ratio $\beta=\omega_c/\omega_p$  between the cyclotron frequency  $\omega_c = QB/m$ and the plasma frequency~$\omega_p$ of the dust grains. 
The wave spectra of unmagnetized and magnetized 2DYP has been previously calculated by theoretical approaches and by computer simulation [2], [3], where, however, the emphasis has been on the primary dispersion branches. These appear in the vicinity of a combination of the plasma frequency $\omega_p$ and the cyclotron frequency $\omega_c$. 
We have solved Eq. (1) by molecular dynamics simulations with integrators specifically adapted to the magnetic field for a system with periodic boundary conditions. From the microscopic particle trajectories one can obtain the wave spectra in the following way: The $k$-space current operator 
\begin{equation}
 \vec j(k, t) = \sum_{j=1}^N \vec v_j(t) \exp[{i k x_j(t)}]. 
\end{equation}
is separated into longitudinal and transverse currents
\begin{equation}
\lambda(k,t) =  \sum_{j=1}^N v_{jx} \exp[{ikx_j}];\label{eq:lambda}
\tau(k,t) =  \sum_{j=1}^N v_{jy} \exp[{ikx_j}],\label{eq:tau}
\end{equation} 
%
%
\begin{figure*}[!t]
 \centering\includegraphics[scale=1.25]{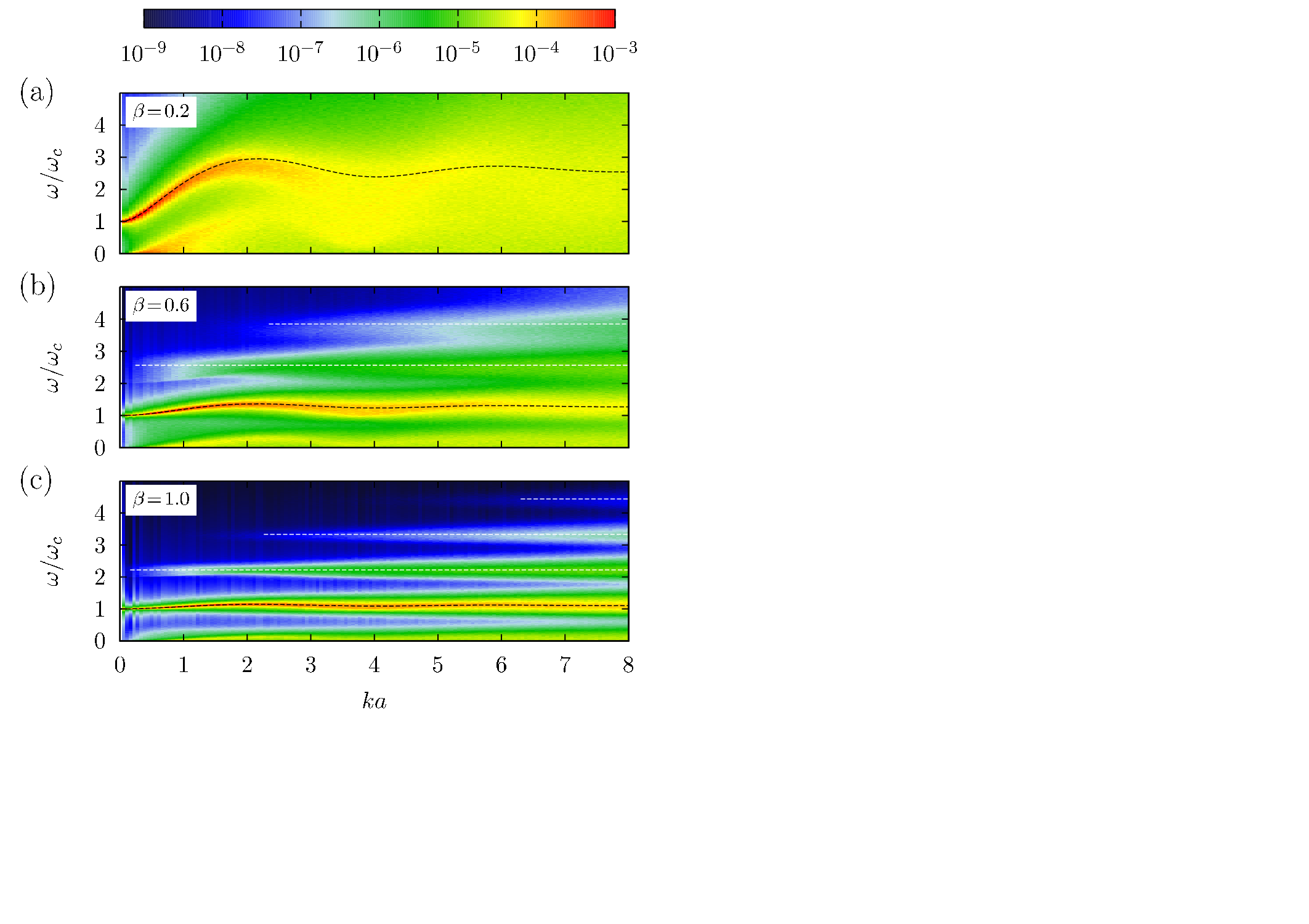}
\caption{The combined collective excitation spectra (color) and the theoretical positions of the dispersion branches (lines) for a two-dimensional Yukawa system with $\kappa=2.0$ and $\Gamma=200$ under the influence of a  magnetic field perpendicular to the particle plane with a strength increasing from (a) to (c).}
\label{fig:geff048}
\end{figure*}
from which the fluctuation spectra are computed by a temporal Fourier transform $\mathcal F_t$: 
\begin{eqnarray}
L(k, \omega) &=& \frac{1}{2\pi N} \lim_{T\rightarrow\infty} \frac{1}{T} \left \vert \mathcal F_t\{\lambda(k, t)\}\right \vert^2, \label{eq:lambdaf} \\
T(k, \omega) &=& \frac{1}{2\pi N} \lim_{T\rightarrow\infty} \frac{1}{T} \left \vert \mathcal F_t\{\tau(k, t)\}\right \vert^2 \label{eq:tauf}.
\end{eqnarray}
The viewgraph shows combined MD collective excitation spectra $L(k,\omega) + T(k,\omega)$. At constant $\kappa=2$ and $\Gamma=200$, a series of increasingly strongly magnetized systems is shown (note that the frequency is scaled by the cyclotron frequency  $\omega_c$). 
In Fig. 1(a), a moderate magnetization of  $\beta=0.2$ is depicted which demonstrates the well-known features of magnetized 2DYP: The two distinct transverse and longitudinal modes from the unmagnetized case are replaced by two modes, the high-frequency magnetoplasmon and the low-frequency magnetoshear mode. The magnetoshear mode is an acoustic mode,  whereas the magnetoplasmon has a $k\rightarrow 0$ limit of $\omega =\omega_c$. 
As we increase the magnetic field [Fig. 1(b-c)] and shift our focus towards the high-frequency part of the spectrum (recall that the scaling factor of the frequency axis is $\omega_c$ which increases linearly with $\beta$), additional dispersion branches appear in the mode spectrum. They enter into the spectrum as a series of higher harmonics and are, in that sense, reminiscent of the classical Bernstein modes. Bernstein modes are, however, spaced by $\omega_c$. It is evident [cf., i.e., Fig. 1(b)], that the high-frequency modes in the strongly coupled magnetized 2DYP are not a pure magnetic effect and are, consequently, spaced by a combination of $\omega_c$  and the Einstein frequency $\omega_E(\Gamma,\kappa)$ [i.e. the average one-particle oscillation frequency in a 2DYP]. We have found an analytic formula for the mode frequencies~[4]  
\begin{eqnarray}
 \omega^2_n(k) & \approx & n^2 \omega^2_{1,\infty}, \qquad \omega^2_{1,\infty} =  \omega_c^2 + 2 \omega_E^2,
\label{eq:om_n}
\end{eqnarray}
which agrees very well the exact spectra, see dotted white lines in the viewgraph, over a broad range of magnetic field strengths [Fig. 1] and all combinations of $\Gamma$ and $\kappa$ [4, 5]. By considering an additional mode damping, we are also able to reproduce the damping of the modes towards small values of $ka$. Recently the existence of these nonlinear modes could be confirmed also for strongly coupled Coulomb systems and for systems subject to noise typical for experiments~[5]. To observe these modes in dusty plasmas it is advantageous to use small (radius below $1\mu m$) particles~[5].

\ifCLASSOPTIONcaptionsoff
  \newpage
\fi

\IEEEtriggeratref{2}

\end{document}